\documentclass[aps,prl,reprint,superscriptaddress,showpacs,preprintnumbers]{revtex4-1}
\usepackage{hyperref}
\usepackage{amsmath,latexsym}
\usepackage{graphicx}
\usepackage{siunitx}
\usepackage{textcomp}
\usepackage{wasysym}
\usepackage{amssymb}
\usepackage{MnSymbol}

\begin{document}
\title{Emergence of Multiscale Dynamics in Colloidal Gels}
\date{\today}
\author{Jae Hyung Cho}
\email[Corresponding author. ]{jaehcho@mit.edu}
\affiliation{Department of Mechanical Engineering, Massachusetts Institute of Technology, Cambridge, Massachusetts 02139, USA}
\author{Roberto Cerbino}
\affiliation{Dipartimento di Biotecnologie Mediche e Medicina Traslazionale, Universit\`a degli Studi di Milano, Via. F.lli Cervi 93, Segrate (MI) I-20090, Italy}
\author{Irmgard Bischofberger}
\affiliation{Department of Mechanical Engineering, Massachusetts Institute of Technology, Cambridge, Massachusetts 02139, USA}

\begin{abstract}
To gain insight into the kinetics of colloidal gel evolution at low particle volume fractions $\phi$, we utilize differential dynamic microscopy to investigate particle aggregation, geometric percolation, and the subsequent transition to nonergodic dynamics. We report the emergence of unexpectedly rich multiscale dynamics upon the onset of nonergodicity, which separates the wave vectors $q$ into three different regimes. In the high-$q$ domain, the gel exhibits $\phi$-independent internal vibrations of fractal clusters. The intermediate-$q$ domain is dominated by density fluctuations at the length scale of the clusters, as evidenced by the $q$ independence of the relaxation time $\tau$. In the low-$q$ domain, the scaling of $\tau$ as $q^{-3}$ suggests that the network appears homogeneous. The transitions between these three regimes introduce two characteristic length scales, distinct from the cluster size.
\end{abstract}


\maketitle

Rich rheological behavior of colloidal gels arises from the coexistence of multiple length and timescales that characterize their structure and dynamics. Through particle aggregation or phase separation, colloidal gels with low particle volume fractions $\phi$ form space-spanning networks of self-similar clusters \cite{Meakin1983,Weitz1984a,Lin1990,Carpineti1992,Lu2008}.  Kinetically arrested, the gels also constantly evolve towards equilibrium via structural rearrangements triggered by thermal agitation and residual stresses \cite{Zaccarelli2007,Zia2014,Gao2015,Cipelletti2000,Manley2005,VanDoorn2017,Bouzid2017}. The ceaseless change in structure, in turn, leads to a continuous evolution of the dynamics. Understanding the microscopic behavior of colloidal gels, therefore, necessitates both spatially and temporally comprehensive investigation. Yet, various scattering techniques used so far to study the dynamics of gel networks have been limited by small ranges of accessible length scales and prolonged data acquisition during which the systems significantly age \cite{Krall1997,Krall1998,Segre2001,Manley2005a,Laurati2009,Romer2000,Scheffold2001,Wyss2001,Liu2013,Calzolari2017,Guo2011,Zhang2017}. This lack of breadth in experimental characterization has prevented a coherent description of the dynamics of colloidal gels. \par

In this Letter, we trace the entire kinetic pathway, from stable suspensions through aged gels, of colloidal gelation and network evolution over large ranges of length and timescales using differential dynamic microscopy (DDM) \cite{Cerbino2008,Giavazzi2009}. The motion of particles and their aggregates initially slows down through two consecutive stages, while the system remains ergodic. As the gel evolves, network fluctuations become greatly suppressed, leading to the onset of nonergodicity. Three dynamically distinct ranges of length scales, or wave vectors $q$, then emerge, unveiling structural hierarchy and macroscopic elasticity of the gel. In the high-$q$ domain, corresponding to length scales considerably smaller than the size of the largest aggregate units, or clusters, the network behaves as internally vibrating fractals. In the intermediate-$q$ domain, the dynamics is dominated by the collective motion over the scale of the clusters. In the low-$q$ domain, the gel fluctuates as a homogeneous elastic network within a viscous solvent. The transitions between these three regimes are determined by both the structure and the elasticity of the network, giving rise to two characteristic length scales. This multiscale dynamics extensively describes colloidal gels from the scale of fractal aggregates to that of a viscoelastic continuum, allowing us to estimate the macroscopic shear modulus in the latter two regimes. \par

This panoramic exploration of the gels throughout their evolution is enabled by DDM that, through optical microscopy, extracts information about the density fluctuations of a sample as in scattering techniques \cite{Cerbino2008}. DDM is less susceptible to the effects of multiple scattering that render the use of traditional far-field scattering techniques impracticable \cite{Giavazzi2014}. Moreover, simultaneous access to ensemble-averaged information at several hundreds of $q$ over more than two decades ($q=0.05 - 10\;\si{\per\micro\meter}$) allows the comprehensive characterization of the evolving samples. We use a CMOS camera (Prime Mono, 2048$\times$2048 pixels, Photometrics) mounted on an inverted microscope (Eclipse TE2000-U, Nikon) with two objectives of magnifications $M=20\times$/60$\times$, and numerical apertures NA = 0.50/1.20 (water immersion), respectively. The samples are loaded in rectangular glass capillary tubes (Vitrocom) of thickness $100\;\si{\micro\meter}$. We repeat some of our experiments with thicker tubes of $200$ and $300\;\si{\micro\meter}$ to check the reproducibility of the results, and confirm no influence of the finite thickness. For the computation of the normalized intermediate scattering function $f(q,{\Delta}t)$, where ${\Delta}t$ denotes the delay time, we use $1000 \;\mathrm{frames}$ acquired at frame rates from $5\;\mathrm{fps}$ through $100\;\mathrm{fps}$. \par

\begin{figure*}[t]
\setlength{\abovecaptionskip}{-25pt}
\hspace*{-0.06cm}\includegraphics[scale=0.11]{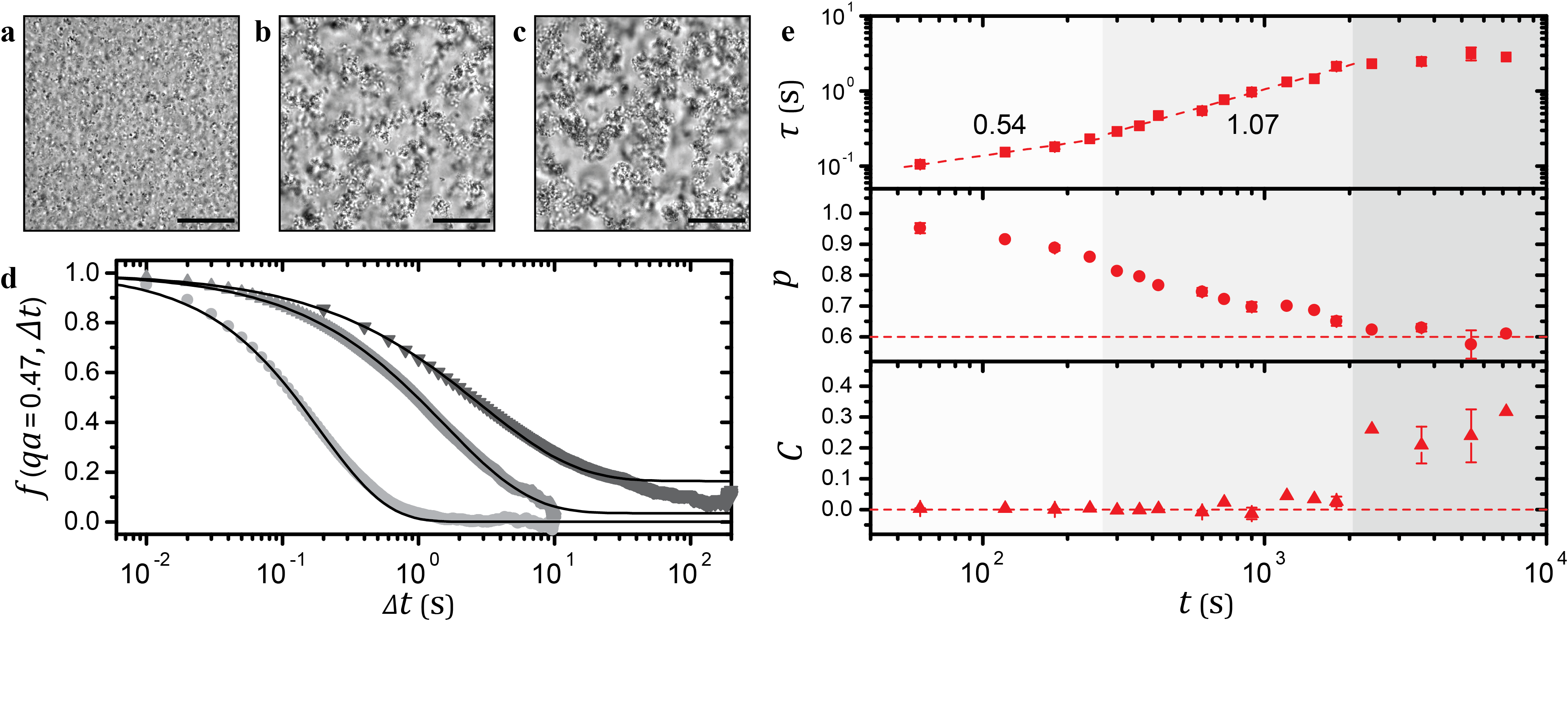}
\caption{\label{fig1} Micrographs showing the temporal evolution of the gel at particle volume fraction $\phi=0.8\%$ at times $t=180\;\si{\s}$ (a), $1500\;\si{\s}$ (b), and $3600\;\si{\s}$ (c) after the onset of aggregation. Scale bars correspond to $30\;\si{\micro\meter}$. (d) Normalized intermediate scattering function $f$ at $t=180\;\si{\s}$ (circle), $1500\;\si{\s}$ (triangle), $3600\;\si{\s}$ (inverted triangle), and corresponding fits (black lines) at a fixed wave vector $q=4.0\;\si{\per\micro\meter}$ or $qa=0.47$, where $a=116.3\;\si{\nano\meter}$ is the hydrodynamic radius of a particle. (e) Temporal evolution of the relaxation time $\tau$, the stretching exponent $p$, and the nonergodicity parameter $C$ at $qa=0.47$. Some error bars are smaller than the symbols.}
\end{figure*}

We utilize polystyrene-poly(N-isopropylacrylamide) (PS-PNIPAM) core-shell colloidal particles synthesized by emulsion polymerization based on the protocol of Ref. \cite{Dingenouts1998,Calzolari2017}. The details of the synthesis are described in the Supplemental Material \cite{[{See the Supplemental Material for the synthesis protocol and the characterization of the PS-PNIPAM particles, the curve-fitting method, and more details of our DDM analysis, which includes Refs. \cite{Giavazzi2017,Meakin1984}}] Supp}. Because of the thin, thermosensitive PNIPAM shell, we can precisely control the initiation of the gelation from a fully stable suspension simply by increasing the system temperature $T$ above the gelation temperature $T_{g}\approx25.5\;\si{\celsius}$. The increase in $T$ reduces the range of the steric repulsion from the shell, allowing the particles to aggregate by van der Waals attraction. To minimize the effect of electrostatic interactions and sedimentation, we add 0.5M of sodium thiocyanate (NaSCN) to screen the charges, and density-match the system using a H\textsubscript{2}O/D\textsubscript{2}O mixture of 52/48 v/v. The hydrodynamic radius $a$ of the particles measured via dynamic light scattering (BI-200SM, Brookhaven Instruments) is $116.3\pm1.8\;\si{\nano\meter}$ at $30\;\si{\celsius}$. \par


We initiate the gelation of a system at $\phi = 0.8\%$ by a sudden temperature increase from $20$ and $30\;\si{\celsius}$ at $t=0\;\si{\s}$. To quantify the dynamics at different $t$ in the micrographs as the ones shown in Figs. \ref{fig1}(a)-\ref{fig1}(c), we focus on the fast dynamics of colloidal gels due to thermal fluctuations \cite{Krall1998}, by assuming the following form of $f(q,{\Delta}t)$
\begin{equation}
f(q,{\Delta}t)=\left[1-C(q)\right]\exp\left[-\left(\frac{{\Delta}t}{{\tau}(q)}\right)^{p(q)}\right]+C(q), \label{eq1}
\end{equation}
which represents a stretched exponential decay from one to the nonergodicity parameter $C(q)$. Here, $\tau(q)$ denotes the relaxation time, and $p(q)$ the stretching exponent. Because of the subsequent relaxation due to slow restructuring of the network \cite{Cipelletti2000}, the large ${\Delta}t$ behavior of $f(q,{\Delta}t)$ often deviates from the fit, as displayed in Fig. \ref{fig1}(d), but this long-time behavior is outside the scope of this work. The details of the fitting procedures are delineated in the Supplemental Material \cite{Supp}. \par  

In the first stage until $t\approx250\;\si{\s}$, the particles form aggregates, as seen in Fig. \ref{fig1}(a). The temporal change of the relaxation time $\tau(q)$ at $q = 4.0\;\si{\per\micro\meter}$ or $qa=0.47$, shown in Fig. \ref{fig1}(e), exhibits a power law with an exponent $0.54\pm0.05$; this growth rate is similar to that of diffusion-limited cluster aggregation (DLCA) \cite{Meakin1983,Weitz1984a}. In DLCA, the mean hydrodynamic radius $R_{h}$ of aggregates scales as $t^{1/d_{f}}=t^{0.57}$ with the fractal dimension $d_{f}=1.75$ \cite{Weitz1984,Weitz1984a,VanDongen1985}. Using $\tau = 1/(Dq^{2})$, where $D$ is the diffusion coefficient of the aggregates, and the Stokes-Einstein relation $D = k_{B}T/(6{\pi}{\eta}R_{h})$, where $k_{B}$ is the Boltzmann constant and $\eta$ the solvent viscosity \cite{Pusey2002}, we can indeed infer that $\tau(t)$ obeys the same power law as  $R_{h}(t)$. Moreover, from the static structure factor $S(q)$ that we calculate based on the squared modulus of Fourier-transformed images \cite{Lu2012,Giavazzi2014}, we find that $d_{f}$ of the aggregates is $1.8\pm0.1$ \cite{Carpineti1992,Supp}, independent of $t$, also in close agreement with DLCA. Although the two relations involving $D$ strictly apply to suspensions of monodisperse particles, the power law exponent of $\tau(t)$ stays nearly independent of $q$, confirming the resemblance of the aggregation to DLCA. \par

The second stage of evolution ($250\;\si{\s}$ $<t<$ $2000\;\si{\s}$) in Fig. \ref{fig1}(e) displays a steeper power law of $\tau(t)$ with an exponent $1.07\pm0.04$, while $p(t)$ monotonically decreases to $\sim0.6$. We suggest that this transition to the second stage is induced by the geometric percolation of the clusters. As the percolation starts to constrain the displacements of the clusters, a marked slowdown of their motion ensues, inducing greater heterogeneity in relaxation times. The stretched exponential relaxation of colloidal gels is generally understood as the superposition of the multiple normal modes subject to overdamped dynamics, each exhibiting a single exponential decay \cite{Krall1998}. In this framework, the distribution of characteristic timescales of the exponentials broadens as $p$ decreases from 1 to approximately 0.65 \cite{Johnston2006}. At all $t$, $p$ is largely independent of $q$ \cite{Supp}. Despite these temporal changes in $\tau$ and $p$, the system retains its ergodicity, which implies that the structural rearrangements after the geometric percolation only gradually give rise to rigidity of the network \cite{Tsurusawa2019a,Zhang2019}.\par

\begin{figure}[t]
\setlength{\abovecaptionskip}{0pt}
\hspace*{-0.03cm}\includegraphics[scale=0.11]{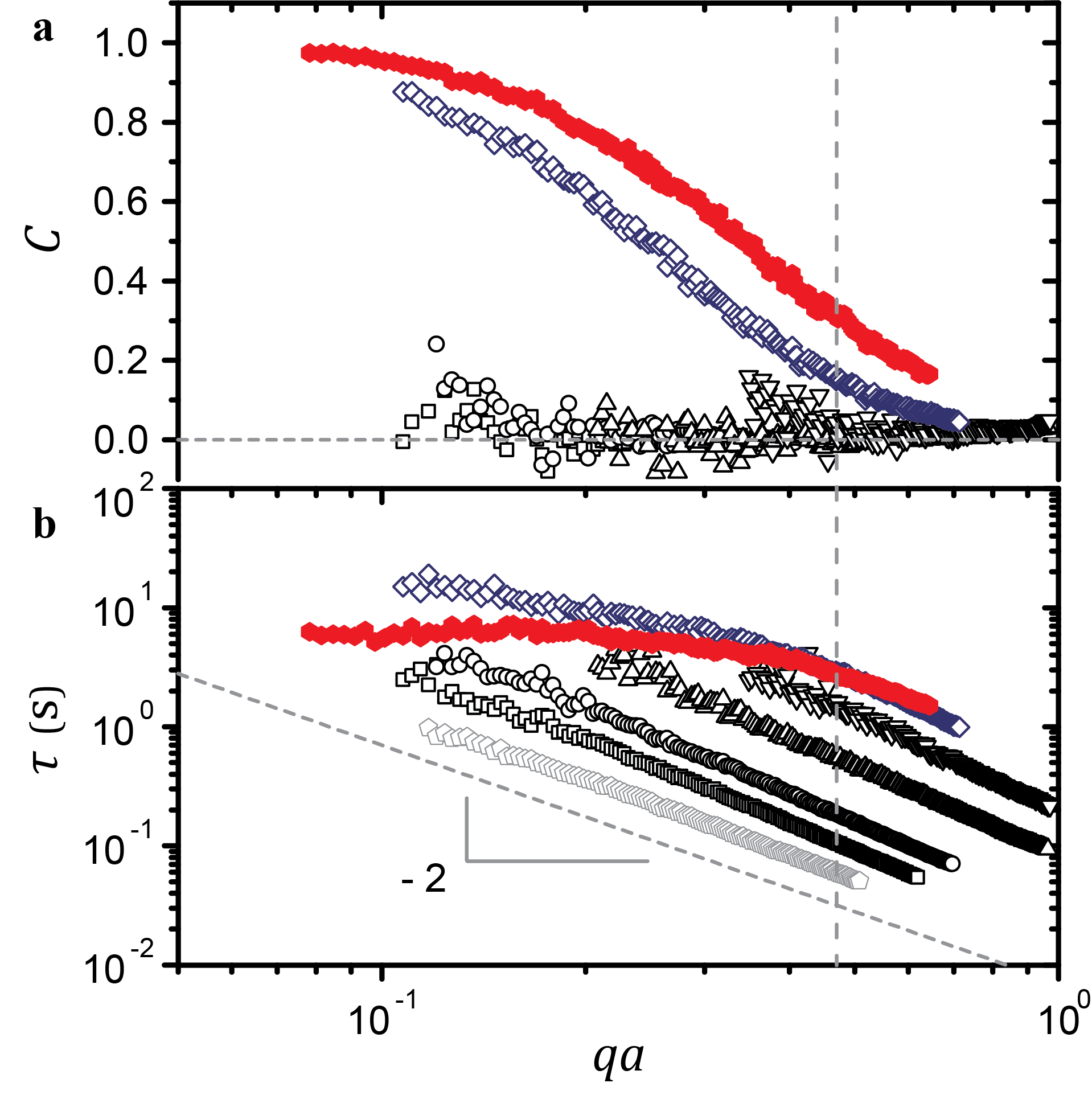}
\caption{\label{fig2} (a) Nonergodicity parameter $C$ and (b) relaxation time $\tau$ of the gel at $\phi = 0.8\%$ as a function of the nondimensionalized wave vector $qa$ at $t=0\;\si{\s}$ (pentagon, stable suspension), $60\;\si{\s}$ (square), $180\;\si{\s}$ (circle), $600\;\si{\s}$ (triangle), $1500\;\si{\s}$ (inverted triangle), $3600\;\si{\s}$ (diamond), and $7200\;\si{\s}$ (filled hexagon). The dashed vertical line denotes $qa=0.47$ corresponding to the data in Fig. \ref{fig1}.}
\end{figure}

After the onset of the third stage ($t\approx2000\;\si{\s}$) shown in Fig. \ref{fig1}(e), the network fluctuations remain partially correlated within the observation time, leading to nonzero values of the nonergodicity parameter $C(q = 4.0\;\si{\per\micro\meter})$. Concurrently, $\tau(q = 4.0\;\si{\per\micro\meter})$ stays nearly constant with $t$. Yet, complex $q$ dependence of the dynamics appears upon this transition. For $t<2000\;\si{\s}$, $C(q)\approx0$ while $\tau(q)\ {\sim}\ q^{-2.2}$, where the power law exponent is close to $-2$ of the dilute suspension of monodisperse particles in Brownian motion \cite{Pusey2002}. For $t>2000\;\si{\s}$, however, $C(q)$ gradually increases with $t$ while monotonically decreasing with $q$, as displayed in Fig. \ref{fig2}(a). Simultaneously, $\tau(q)$ flattens, and eventually becomes independent of $q$ for $qa<0.23$, as shown in Fig. \ref{fig2}(b). This flattening of $\tau(q)$ marks the end of major temporal evolution, after which all the parameters remain nearly unchanged. \par

\begin{figure}[t]
\setlength{\abovecaptionskip}{0pt}
\hspace*{-0.03cm}\includegraphics[scale=0.11]{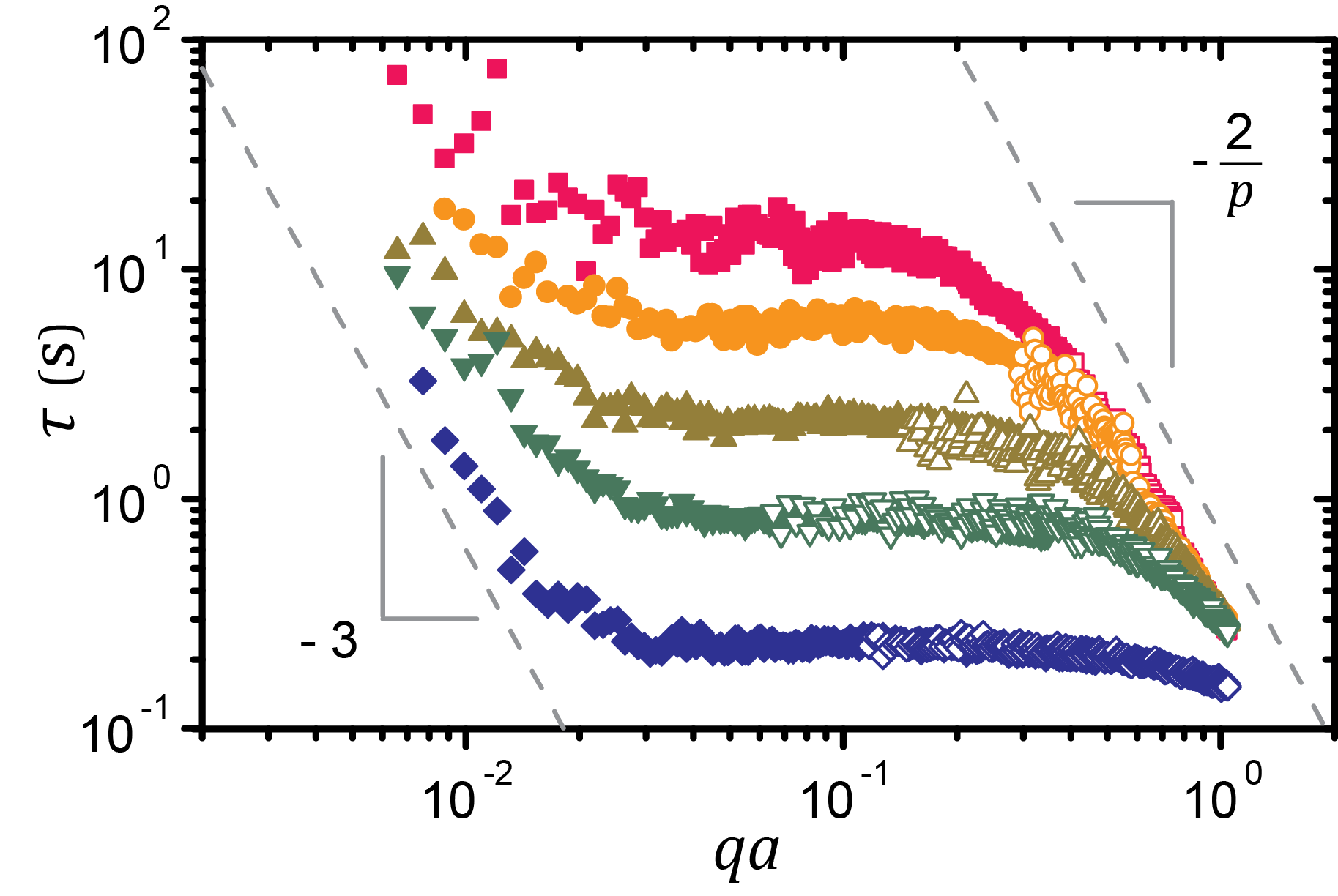}
\caption{\label{fig3} Relaxation time $\tau$ of the aged gels in quasisteady states as a function of the nondimensionalized wave vector $qa$ at $\phi=0.5\%$ (square), $0.8\%$ (circle), $1.0\%$ (triangle), $1.5\%$ (inverted triangle), and $2.0\%$ (diamond). Filled symbols denote data obtained with a 20$\times$ objective and open symbols with a 60$\times$ objective.}
\end{figure}

For a more comprehensive inspection of the $q$ dependence of the dynamics in the aged gels, we employ two objectives $(M=20\times$ and $60\times)$ to extract $\tau(q)$ of the gels at five different $\phi$ $(= 0.5$, 0.8, 1.0, 1.5, and $2.0\%)$ in their quasisteady states at sufficiently large $t$. For each $\phi$, three distinct regimes of dynamics emerge, as displayed in Fig. \ref{fig3}. In the remainder of this Letter, we show how this transition into nonergodic, multiscale dynamics reveals the structure and the macroscopic elasticity of the gels. \par

In the high-$q$ domain, we observe dynamical hallmarks of fractals. The averaged internal structure of fractal aggregates is fully defined by $d_{f}$ only, and indeed, $\tau(q)$ of all $\phi$ asymptotically collapse onto a single line, indicating the presence of $\phi$-independent structures at the smallest length scales probed. Furthermore, according to the model proposed by Reuveni \textit{et al}. \cite{Reuveni2012,Reuveni2012a}, the internal dynamics of a vibrating fractal under thermal perturbation and strong viscous damping for $qR_{g}\gg1$, where $R_{g}$ is the cluster radius of gyration, obeys the following scaling relation in the absence of translation and rotation:
\begin{equation}
\tau\ \sim\ q^{-2/p}. \label{eq2}
\end{equation}
We obtain $p=0.66\pm0.02$ for all $\phi$ \cite{Supp}, which yields the value of $-2/p$ consistent with our high-$q$ power law exponent. \par 

In the intermediate-$q$ domain, where $\tau$ is independent of $q$, the motion at the length scale of clusters dominates the gel dynamics. In the model developed by Krall and Weitz for fractal gels \cite{Krall1997,Krall1998}, $f(q,{\Delta}t)$ is determined by $\tau$ and the maximum mean squared displacement $\delta^{2}$ at the length scale of the clusters as follows:
\begin{equation}
f(q,{\Delta}t) = \exp\left(-\frac{q^{2}\delta^{2}}{4}\left\{1-\exp\left[-{\left(\frac{{\Delta}t}{\tau}\right)}^{p}\right]\right\}\right), \label{eq3}
\end{equation}
which, for $q^{2}\delta^{2}\ll1$, can be simplified to
\begin{equation}
f(q,{\Delta}t) \approx \frac{q^{2}\delta^{2}}{4}\exp\left[-{\left(\frac{{\Delta}t}{\tau}\right)}^{p}\right]+\left(1-\frac{q^{2}\delta^{2}}{4}\right), \label{eq4}
\end{equation}
an equivalent functional form to our model in Eq. (\ref{eq1}). From Eqs. (\ref{eq1}) and (\ref{eq3}) in the limit of ${\Delta}t \rightarrow \infty$,
\begin{equation}
f(q,{\Delta}t \rightarrow \infty) = \exp\left(-\frac{q^{2}\delta^{2}}{4}\right) = C(q). \label{eq5}
\end{equation}
The resulting $\delta^{2}$ from curve fits to $C(q)$ satisfies $q^{2}\delta^{2}\ll1$ for the intermediate-$q$ domain \cite{Supp}. Provided that the microscopic elasticity of the gels is governed by the local spring constant $\kappa \sim (a/\xi)^{\beta}$ between two particles of radius $a$ separated by the distance $\xi$ within a fractal cluster, where $\beta$ is the elasticity exponent \cite{Shih1990,DeRooij1994,West1994}, $\tau$ and $\delta^{2}$ scale with $\phi$ as
\begin{equation}
\tau\; {\sim}\; \phi^{-\left(\beta+1\right)/\left(3-d_{f}\right)},\quad \delta^{2}\; \sim\; \phi^{-\beta/\left(3-d_{f}\right)}, \label{eq6}
\end{equation}
respectively \cite{Krall1998}. We measure the mean value of the $q$-independent relaxation time $\tau_{m}$, and observe the scalings of $\tau_{m} \sim \phi^{-2.93\pm0.15}$ and $\delta^{2} \sim \phi^{-2.07\pm0.04}$. With $d_{f}=1.8$, consistent values of $\beta=2.52\pm0.18$ and $2.48\pm0.05$, respectively, are obtained, which reflect considerable structural rearrangements during the network formation \cite{DeRooij1994,Kantor1984,Romer2014,Supp}. \par

\begin{figure}[t]
\setlength{\abovecaptionskip}{-15pt}
\hspace*{-0.03cm}\includegraphics[scale=0.11]{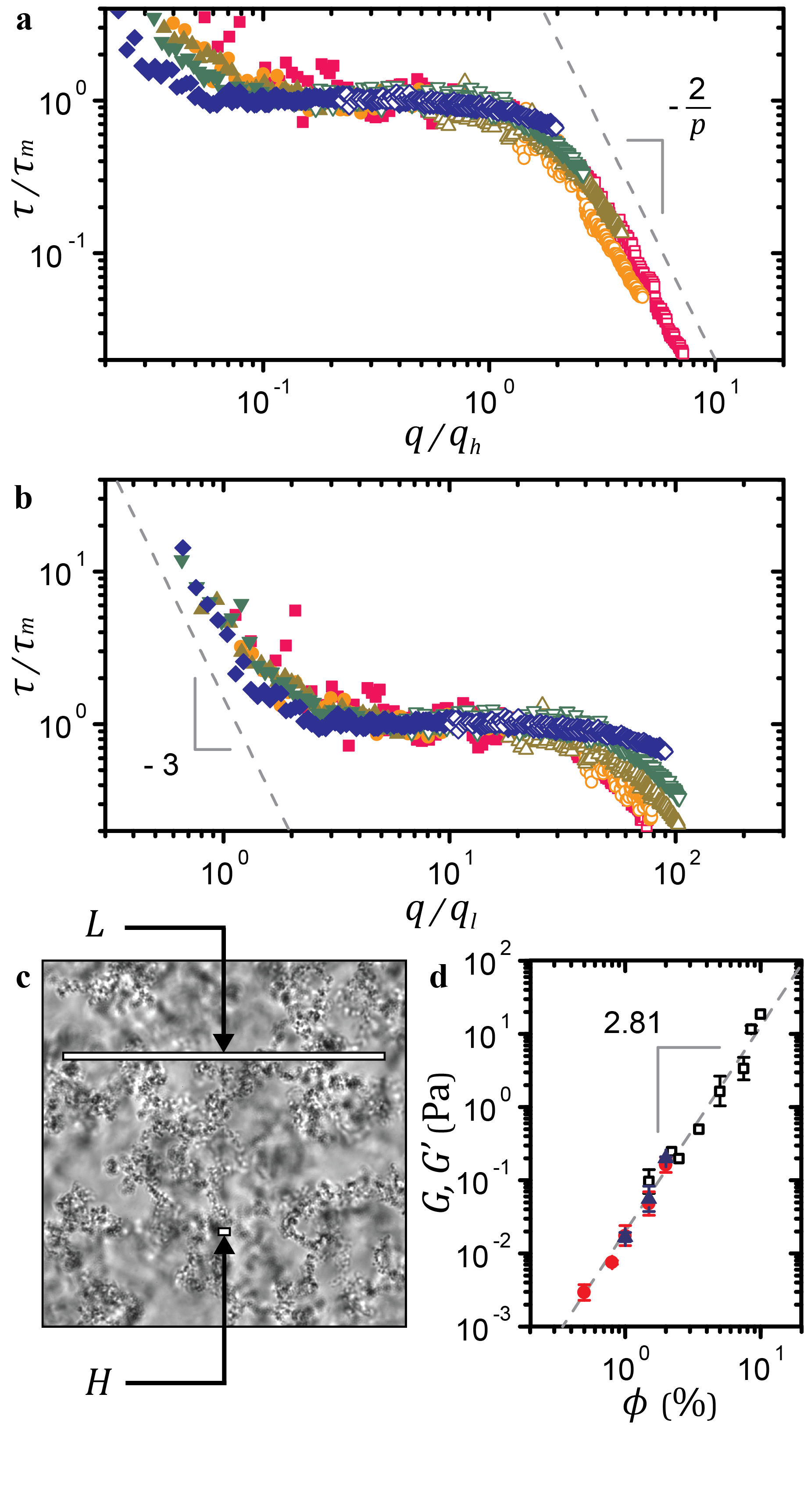}
\caption{\label{fig4} Relaxation time $\tau$ of all $\phi$ scaled with the wave vectors at the intermediate-to-high-$q$ transition $q_{h}$ (a) or the low-to-intermediate-$q$ transition $q_{l}$ (b) in the abscissa and the mean $\tau$ of the intermediate-$q$ domain $\tau_{m}$ in the ordinate. (c) Micrograph of the aged gel at $\phi=0.8\%$. Scale bars indicate the transition length scales $H\ =2\pi/q_{h}\ \approx\ 3.3\;\si{\micro\meter}$, and $L\ =2\pi/q_{l}\ \approx\ 99.4\;\si{\micro\meter}$. (d) Storage modulus $G'$ (square) obtained from conventional rheometry and shear modulus $G$ estimated from DDM in the intermediate-$q$ (filled circle) and the low-$q$ (filled triangle) domains. The DDM estimates show consistent power laws.}
\end{figure}

We demonstrate that our model simultaneously exhibits the dynamics described in the two models by Reuveni \textit{et al}. \cite{Reuveni2012a} and by Krall and Weitz \cite{Krall1998}, as it separates the $q$ dependence of the characteristic relaxation time and that of the nonergodicity into two parameters $\tau(q)$ and $C(q)$, respectively, in Eq. (\ref{eq1}). The anomalous diffusion of the subcluster aggregates is ergodic for $qR_{g}\gg1$ \cite{Reuveni2012a}, while the fluctuations at the length scale of the clusters are nonergodic \cite{Krall1998}. Yet, the $q^{2}$ term in Eq. (\ref{eq3}) determines the $q$ dependence of both the timescale of the decay and the plateau of $f$ as ${\Delta}t\rightarrow \infty$, by assuming nonergodic processes at all $q$ \cite{Krall1998,Supp}. In our model, $\tau(q)$ and $C(q)$ allow independent determination of within what time and how far, respectively, scatterers move at each $q$. We thus suggest that, at the transition between the intermediate-$q$ and the high-$q$ regimes, the relaxation time of the internal vibrations captured in Eq. (\ref{eq1}) simply scales as $\tau_{m}$, which from Eqs. (\ref{eq2}) and (\ref{eq6}) leads to
\begin{equation}
q_{h}\ \sim\ \phi^{\left[p\left(\beta+1\right)\right]/\left[2\left(3-d_{f}\right)\right]}, \label{eq7}
\end{equation}
where $q_{h}$ denotes the intermediate-to-high-$q$ transition wave vector. With $p=0.66$ and $\beta = 2.50$, Eq. (\ref{eq7}) yields $q_{h} \sim \phi^{0.96\pm0.07}$. Using $q_{h}$ and $\tau_{m}$ to scale $q$ and $\tau(q)$, we find that our data of all $\phi$ collapse onto a master curve in the plateau and the high-$q$ domain, as shown in Fig. \ref{fig4}(a). \par

In the low-$q$ regime, the gels display overdamped dynamics characteristic of a homogeneous viscoelastic medium. The effective spring constant $\kappa$ between two particles is independent of their distance $\xi$, if $\xi$ is greater than the smallest length scale $L$ at which the continuum assumption holds true \cite{Dinsmore2006}. The magnitude of the frictional force that a scatterer experiences in a homogeneous two-phase continuum, however, scales as its volume $V$ \cite{Levine2000,Levine2001}, as every differential volume undergoes the same amount of viscous coupling between the two phases. Because of the tenuous and porous structure of the network, the total frictional force is expected to be proportional to the number of constituent particles $N$, which in a continuum scales as $N \sim {\phi}V$. Thus, in the low-$q$ domain,
\begin{equation}
\tau\  \sim\ \frac{{\eta}a \left({\phi}V/a^{3}\right)}{GL}\ \sim\ \frac{{\eta}{\phi}q_{l}}{Ga^{2}}q^{-3}, \label{eq8}
\end{equation}
where $G \sim \kappa/L$ denotes the macroscopic shear modulus, with $L$ chosen as the characteristic length scale of the elastic forces and $q_{l}=2\pi/L$. Indeed, $\tau(q)$ at $\phi=2.0\%$ exhibits the scaling of $\tau \sim q^{-3}$ for $qa<0.012$, as shown in Fig. \ref{fig3}. For Eq. (\ref{eq8}) to be valid at all $\phi$, we expect $\tau(q)={\psi}(\phi)q^{-3}$, where ${\psi}(\phi) \sim {\tau}_{m}(\phi){q_{l}(\phi)}^3$. Because $\tau_{m}$ denotes the timescale of the floppiest mode of the gel network, it also sets $G$ by
\begin{equation}
G=\frac{6\pi{\eta}b_{m}}{\tau_{m}}, \label{eq9}
\end{equation}
where $b_{m}$ is a correction factor \cite{Krall1998}. Substituting the general form of $\tau(q)$ into Eq. (\ref{eq8}) and using $\tau_{m}(\phi)\sim{G(\phi)}^{-1}$ from Eq. (\ref{eq9}), therefore, results in
\begin{equation}
q_{l} \sim {\phi}^{0.5}. \label{eq10}
\end{equation}
Scaling $\tau(q)$ of all $\phi$ with the resulting $q_{l}$ leads to a master curve in the low-$q$ and the intermediate-$q$ domains, as shown in Fig. \ref{fig4}(b). The power law in Eq. (\ref{eq10}) differs from that of the DLCA cluster radius ${R_{c}}^{-1}\sim{\phi}^{1/(3-d_{f})}={\phi}^{0.8}$ \cite{Shih1990}, since the macroscopic elasticity, as well as the structure, governs the dynamical length scale $L$. We insert scale bars of length $L=2\pi/q_{l}$ and $H=2\pi/q_{h}$ in Fig. \ref{fig4}(c) at $\phi=0.8\%$, to visually highlight the contrast between the length scales of the low-$q$ and the high-$q$ regimes. \par

The direct link between $\tau$ and $G$ indicates that we can estimate the macroscopic shear modulus $G$ of the gels by measuring the microscopic relaxation time $\tau$ in either the intermediate-$q$ domain or, as long as we can identify $q_{l}$, the low-$q$ domain. Using Eq. (\ref{eq9}) with $b_{m}=2.8$, we observe that the resulting values of $G$ yield a smooth continuation of the storage modulus $G'$ obtained from conventional rheometry (AR-G2, TA Instruments), as shown in Fig. \ref{fig4}(d). Alternatively, $G$ can be estimated by measuring $\psi$ and $q_{l}$ from Fig. \ref{fig3}, if the power law behavior of $\tau \sim q^{-3}$ is accessible. The following rearranged form of Eq. (\ref{eq8})
\begin{equation}
G=\frac{\eta{\phi}q_{l}b_{l}}{a^{2}\psi}, \label{eq11}
\end{equation}
where $b_{l}=1.2$ is a correction factor, yields consistent scaling behavior of $G$, as displayed in Fig. \ref{fig4}(d). \par

Our dynamical investigation of evolving colloidal gels unveils the extensive kinetic route through particle aggregation, geometric percolation, and the emergence of nonergodicity that establishes distinct dynamical regimes at different length scales. In particular, we show that internal vibrations of random fractals, cluster-dominated fluctuations, and the homogeneity of a viscoelastic medium simultaneously define the nonergodic colloidal gels. Consequently, our results not only demonstrate links among different models, but clarify their limits. We expect similar applications of DDM to other types of soft matter, such as biopolymer particle-network mixtures \cite{Wulstein2019,Burla2020} or dense suspensions of active particles \cite{Janssen2019,Berthier2019}, can likewise provide comprehensive descriptions of their nonergodic, multiscale dynamics \cite{Cerbino2017}. \par

\begin{acknowledgments}
We thank Veronique Trappe and Fabio Giavazzi for helpful discussions. J.H.C. and I.B. acknowledge support from the MIT Research Support Committee and Kwanjeong Educational Foundation, Awards No. 16AmB02M and No. 18AmB59D. R.C. acknowledges support from Regione Lombardia and CARIPLO Foundation, Grant No. 2016-0998.
\end{acknowledgments}

\bibliographystyle{apsrev4-1}
\bibliography{DDM_paper_Aug_19_mod.bib}

\end{document}


\title{Emergence of Multiscale Dynamics in Colloidal Gels: Supplemental Material}
\date{\today}
\author{Jae Hyung Cho}
\affiliation{Department of Mechanical Engineering, Massachusetts Institute of Technology, Cambridge, Massachusetts 02139, USA}
\author{Roberto Cerbino}
\affiliation{Dipartimento di Biotecnologie Mediche e Medicina Traslazionale, Universit\`a degli Studi di Milano, Via. F.lli Cervi 93, Segrate (MI) I-20090, Italy}
\author{Irmgard Bischofberger}
\affiliation{Department of Mechanical Engineering, Massachusetts Institute of Technology, Cambridge, Massachusetts 02139, USA}

\maketitle

\subsection{1. Particle synthesis and characterization}
For the first part of the synthesis of the polystyrene-poly(N-isopropylacrylamide) (PS-PNIPAM) core-shell particles, we follow the protocol described in Ref. \cite{Calzolari2017}, which is slightly modified from that of Ref. \cite{Dingenouts1998}. In a $1\;\si{\liter}$ flask equipped with a stirrer, a reflex condenser, and a gas inlet, $25.02\;\si{\gram}$ of N-isopropylacrylamide (NIPAM, Acros Organics) and $0.2008\;\si{\gram}$ of the stabilizer sodium dodecyl sulfate (SDS, Sigma-Aldrich) are dissolved in $525.14\;\si{\gram}$ of DI water. After the solution is bubbled with nitrogen for $30\;\si{\min}$, $142.75\;\si{\gram}$ of styrene (Sigma-Aldrich) is added, and the mixture is heated to $80\si{\celsius}$ in nitrogen atmosphere. Then $0.3521\;\si{\gram}$ of the initiator potassium persulfate (KPS, Acros Organics) dissolved in $15.00\;\si{\gram}$ of DI water is added to the mixture. After $6\;\si{\hour}$, the dispersion is cooled to room temperature and cleaned through repeated centrifugation and supernatant exchange. \par

For the second part, we use the seeded emulsion polymerization in Ref. \cite{Dingenouts1998} that increases the thickness of the PNIPAM shell, with slight modification of the ratio of the materials. For each $100\;\si{\gram}$ of the particles obtained from the first part, $12.58\;\si{\gram}$ of NIPAM and $0.8994\;\si{\gram}$ of the crosslinker N,N'-methylenebis(acrylamide) (BIS, Sigma-Aldrich) are added, and the mixture is heated to $80\si{\celsius}$. After the addition of $0.1264\;\si{\gram}$ of KPS dissolved in $9.43\;\si{\gram}$ of DI water, the mixture is stirred for $4\;\si{\hour}$. The suspension is then cooled to room temperature, and cleaned by dialysis against DI water for approximately 4 weeks. \par 

We add sodium thiocyanate (NaSCN) to screen the charges of the synthesized particles. To probe the amount of NaSCN necessary to neglect the range of electrostatic repulsions, we gradually increase the concentration $c$ of NaSCN and measure the viscoelastic moduli of the resulting gels at the temperature $T=30\si{\celsius}$. We observe that the moduli gradually increase with $c$, until they remain constant in the range of $c=0.3-0.7\;\si{\molar}$. Hence we conduct all of our experiments at $c=0.5\;\si{\molar}$. \par

We estimate the value of the gelation temperature $T_{g}$ by measuring $T$, at which the storage modulus $G'$ becomes larger than the loss modulus $G''$, during a temperature ramp experiment at the ramp rate of $0.2\;\si{\celsius\per\min}$ that is sufficiently slow to ensure uniform sample temperature \cite{Calzolari2017}. We estimate the interparticle attraction strength at $T=30\si{\celsius}$ to be $\sim3.5\;k_{B}T$, where $k_{B}$ is the Boltzmann constant, from the dependence of $G'$ on the particle volume fraction $\phi$, using the van der Waals potential \cite{Calzolari2017}. Yet, we expect strong influence of noncentral forces between neighboring particles on the rheological properties of our gels, which would effectively lead to a larger attraction strength.  \par

We measure the radius of gyration $r_{g}$ of a particle to be $81.4\pm0.7\;\si{\nano\meter}$ at $30\si{\celsius}$ and $82.3\pm0.9\;\si{\nano\meter}$ at $20\si{\celsius}$ by using the Guinier plot via static light scattering, which corresponds to a hard sphere radius $r=\sqrt{5/3}r_{g}$ of $r=105.1\pm1.0\;\si{\nano\meter}$ at $30\si{\celsius}$ and $r=106.2\pm1.2\;\si{\nano\meter}$ at $20\si{\celsius}$. In addition, we measure the hydrodynamic radius $a$ to be $116.3\pm1.8\;\si{\nano\meter}$ at $30\si{\celsius}$ and $139.9\pm1.7\;\si{\nano\meter}$ at $20\si{\celsius}$ by using the Stokes-Einstein relation via dynamic light scattering. Due to the sparse, brush-like structure of the PNIPAM shell \cite{Dingenouts1998}, $r$ is expected to represent the approximate radius of the dense PS core, as evidenced by the small $T$-dependence of $r$. The hydrodynamic radius $a$ is expected to overestimate the actual radius of the entire core-shell particle, as the Stokes drag assumes a perfect, smooth sphere. We thus infer a negligible thickness of the PNIPAM shell compared to the size of the PS core, and use the density of polystyrene $\rho=1.05\;\si{\gram\per\cm^{3}}$ to estimate $\phi$ from the mass change of a sample after drying it in an oven. \par

\subsection{2. Rheological measurements}
For the measurement of the storage modulus $G'$, we use an AR-G2 rheometer (TA Instruments) to perform small amplitude oscillatory shear experiments. For each sample, upon a sudden temperature increase from $20\si{\celsius}$ to $30\si{\celsius}$ at time $t=0\;\si{\s}$, we monitor the temporal evolution of the viscoelastic moduli during a time sweep. When the change in the moduli becomes very slow at $t\approx2400\;\si{\s}$, an amplitude sweep is followed (for samples at low $\phi$, the time sweep is prolonged until $t\approx4200\;\si{\s}$ to account for the slower kinetics). The values of $G'$ reported are the mean values within the linear range (strain amplitude ${\gamma}_{0}\approx0.001-0.1$), where the moduli are constant. We use a $40\;\si{\milli\meter}$ cone-plate geometry, and all the experiments are performed at the frequency $f=1\;\si{\hertz}$. \par

\subsection{3. Curve-fitting method}
The image structure function $D(q,\Delta t)$ computed with the DDM algorithm can be expressed as
\begin{equation}
D(q,{\Delta}t)=A(q)\left(1-f(q,{\Delta}t)\right)+B(q), \label{eq1}
\end{equation}
where $A(q)$ is determined by optical properties of the microscope and static information about the sample, and $B(q)$ represents the level of the camera noise \cite{Giavazzi2014}. The form of the normalized intermediate scattering function we assume is
\begin{equation}
f(q,{\Delta}t)=\left(1-C(q)\right)\exp\left[-\left(\frac{{\Delta}t}{{\tau}(q)}\right)^{p(q)}\right]+C(q), \label{eq2}
\end{equation}
where $C(q)$ is the nonergodicity parameter, ${\Delta}t$ the delay time, $\tau(q)$ the relaxation time, and $p(q)$ the stretching exponent. We determine $A(q)$ and $B(q)$ prior to the curve-fitting. The value of $B(q)$ is independent of the wave vector $q$ if the detection noise of the camera is uncorrelated in space and time \cite{Giavazzi2017}. Indeed we observe that $D(q,{\Delta}t)$  becomes independent of both $q$ and ${\Delta}t$ in the highest-$q$ domain accessible ($q>9.5\;\si{\per\micro\meter}$ for the $20\times$ objective and $q>28\;\si{\per\micro\meter}$ for the $60\times$ objective), where $A(q)$ approaches zero. We calculate the mean of $D(q,{\Delta}t)$ in this domain and equate the resulting value to $B(q)$. \par

In linear space invariant imaging, where the sample density field is linearly mapped onto the image intensity field, the ensemble-averaged squared modulus of Fourier-transformed images can be expressed as
\begin{equation}
\left<{\abs{\hat{i}(q)}}^{2}\right>_{E} \simeq \frac{A(q)}{2}+\frac{B(q)}{2}, \label{eq3}
\end{equation}
provided that the non-ideal contributions arising from imperfections, such as scratches, stains, or dust particles, along the optical path are negligible compared to those from the sample \cite{Giavazzi2014}. Hence we calculate $A(q)$ from $\left<{\abs{\hat{i}(q)}}^{2}\right>_{E}$ and $B(q)$. \par

\begin{figure*}[t]
\setlength{\abovecaptionskip}{-25pt}
\hspace*{-0.03cm}\includegraphics[scale=0.12]{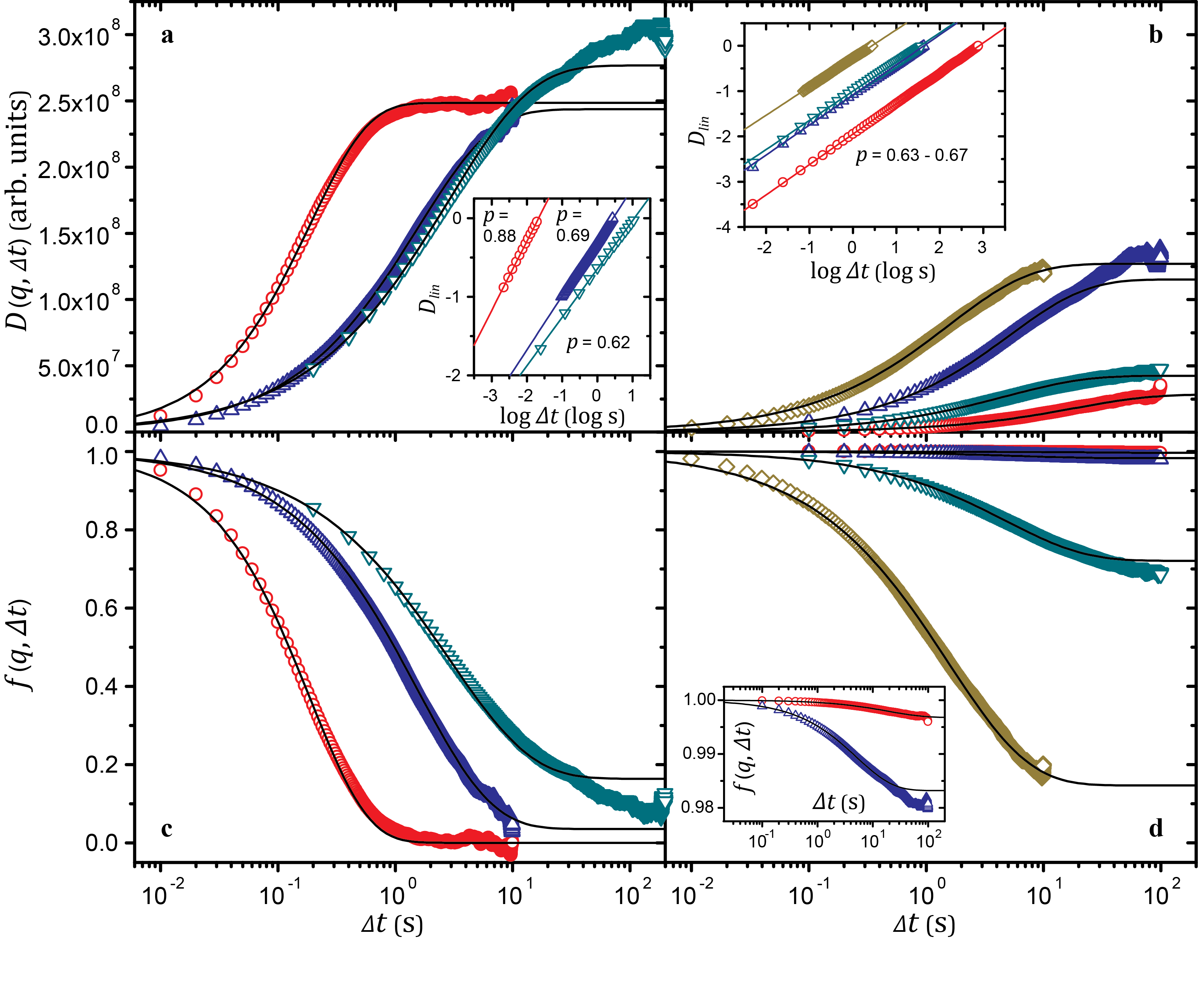}
\caption{\label{supp_fig1} Image structure function $D(q,{\Delta}t)$ (a, b), normalized intermediate scattering function $f(q,{\Delta}t)$ (c, d), and the fitting curves (black lines) for the system at particle volume fraction $\phi=0.8\%$. The correlation functions in (a) and (c) correspond to the sample ages $t=180\;\si{\s}$ ($\medcircle$), $1500\;\si{\s}$ ($\triangle$), and $3600\;\si{\s}$ ($\bigtriangledown$) at the wave vector $q=4.0\;\si{\per\micro\meter}$ or $qa=0.47$, where $a$ is the hydrodynamic radius of a particle. The correlation functions in (b) and (d) correspond to $qa=0.0088$ ($\medcircle$), $qa=0.058$ ($\triangle$), $qa=0.23$ ($\bigtriangledown$), and $qa=0.58$ ($\Diamond$) of the aged gel. Insets (a,b): linearized image structure function $D_{lin}(\log{\Delta}t)$, whose slope equals the stretching exponent $p$. Inset (d): a zoomed-in image of the curves for $qa=0.0088$ ($\medcircle$) and $qa=0.058$ ($\triangle$).}
\end{figure*}

We use \textit{lsqcurvefit} in MATLAB to extract the three remaining parameters from curve-fits. We first plot $D(q,{\Delta}t)$ as a function of ${\Delta}t$ in linear-log scales to estimate the time window ${\Delta}t=[0, {\Delta}t^{*}$] of the fast relaxation dynamics by ensuring that a point of inflection is captured in the domain. Such a point of inflection indicates that the relaxation time $\tau(q) \lesssim {\Delta}t^{*}$, and therefore the plateau of $D(q,{\Delta}t\rightarrow \infty)=A(q)\left(1-C(q)\right)+B(q)$, is properly captured by fitting the model to the data. The nonergodicity parameter $C(q)$ is obtained from this plateau value. To improve the reliability of the fitting values of the relaxation time $\tau(q)$ and the stretching exponent $p(q)$, we then linearize Eq. (\ref{eq1}) with Eq. (\ref{eq2}) by plotting $D_{lin}=\log \left( -\log \left[ 1-\frac{D(q,{\Delta}t)-B(q)}{A(q)(1-C(q))} \right] \right)$ as a function of $\log({\Delta}t)$ such that the slope is equal to $p(q)$ and the $y$-intercept $-p(q)\log(\tau(q))$. The resulting plot is linear for early $\log({\Delta}t)$ and typically starts to show a gradually decreasing slope for later $\log({\Delta}t)$. By performing linear regression to the data in the early times as shown in the insets of Fig. \ref{supp_fig1}(a,b), we re-evaluate $\tau(q)$ and $p(q)$. We then use these values to plot Eq. (\ref{eq1}) and check the quality of the resulting fit. We find that this last step of correction is crucial, as $p$ is highly sensitive to the fitting domain of ${\Delta}t$. The sensitivity originates from the properties of the stretched exponential function, which inherently contains a broad distribution of constituent time scales \cite{Johnston2006}. Examples of the final fits to $D(q,{\Delta}t)$ and the corresponding $f(q,{\Delta}t)$ at the particle volume fraction $\phi=0.8\%$ are displayed in Fig. \ref{supp_fig1}. As the sample age $t$ since the onset of aggregation increases, the curves become more stretched, while the relaxation time increases, as shown in Fig. \ref{supp_fig1}(c) (same data as in Fig. 1(d) of the main text). For the aged gels, the nonergodicity parameter, as visualized by the large ${\Delta}t$ plateau in $f$, decreases with $q$, as displayed in Fig. \ref{supp_fig1}(d). \par

\subsection{4. Measurement of the fractal dimension $d_{f}$}

\begin{figure}[t]
\setlength{\abovecaptionskip}{0pt}
\hspace*{-0.03cm}\includegraphics[scale=0.12]{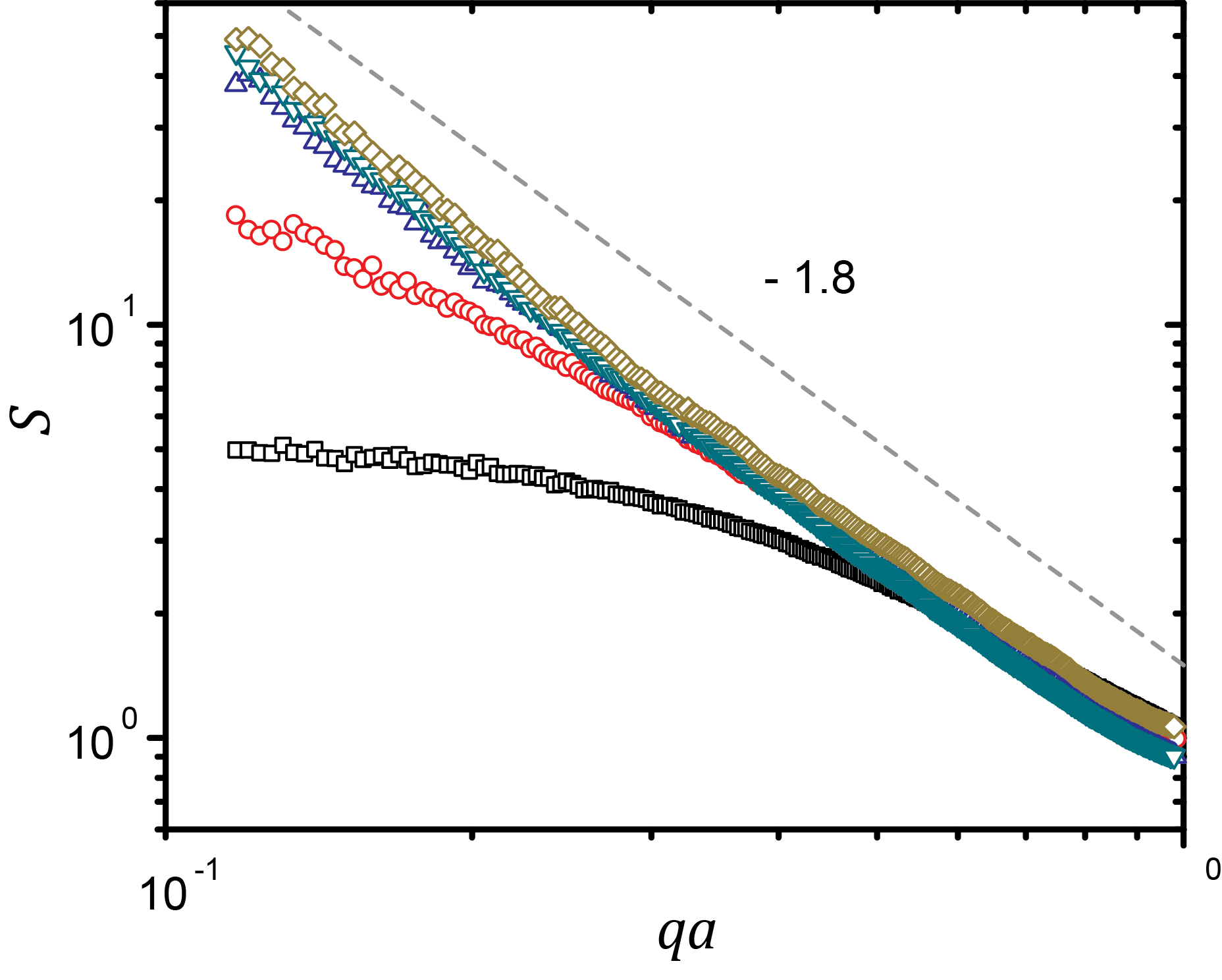}
\caption{\label{supp_fig2} Static structure factor $S$ of the gel at $\phi=0.8\%$ as a function of the nondimensionalized wave vector $qa$ at times $t=60$ \si{\s} ($\square$), 180 \si{\s} ($\medcircle$), 600 \si{\s} ($\triangle$), 900 \si{\s} ($\bigtriangledown$), and 1800 \si{\s} ($\Diamond$) after the initiation of gelation.}
\end{figure}

The static structure factor $S(q)$ is calculated by dividing the squared modulus of Fourier-transformed images of a gel by that of the same system before the initiation of gelation for each $q$ \cite{Giavazzi2014,Lu2012}. The resulting $S(q)$ in the high-$q$ regime displays a power law with an exponent $-1.8\pm0.1$, independent of time $t$, as shown in Fig. \ref{supp_fig2} for the particle volume fraction $\phi=0.8\%$. We confirm that the exponent is independent of $\phi$ as well. The absolute value of the exponent represents the fractal dimension $d_{f} \approx 1.8\pm0.1$ \cite{Carpineti1992}. \par

At lower $q$, a peak appears in $S(q)$, which typically indicates the uniformity of the cluster size \cite{Carpineti1992}. However, we find that $S(q)$ from our experiments is significantly influenced in the low-$q$ domain by the optical properties of the experimental setup. Hence we use $S(q)$ only to determine $d_{f}$ in the high-$q$ domain. \par

\subsection{5. $q$-independence of $p(q)$}

\begin{figure}[h]
\setlength{\abovecaptionskip}{0pt}
\hspace*{-0.03cm}\includegraphics[scale=0.12]{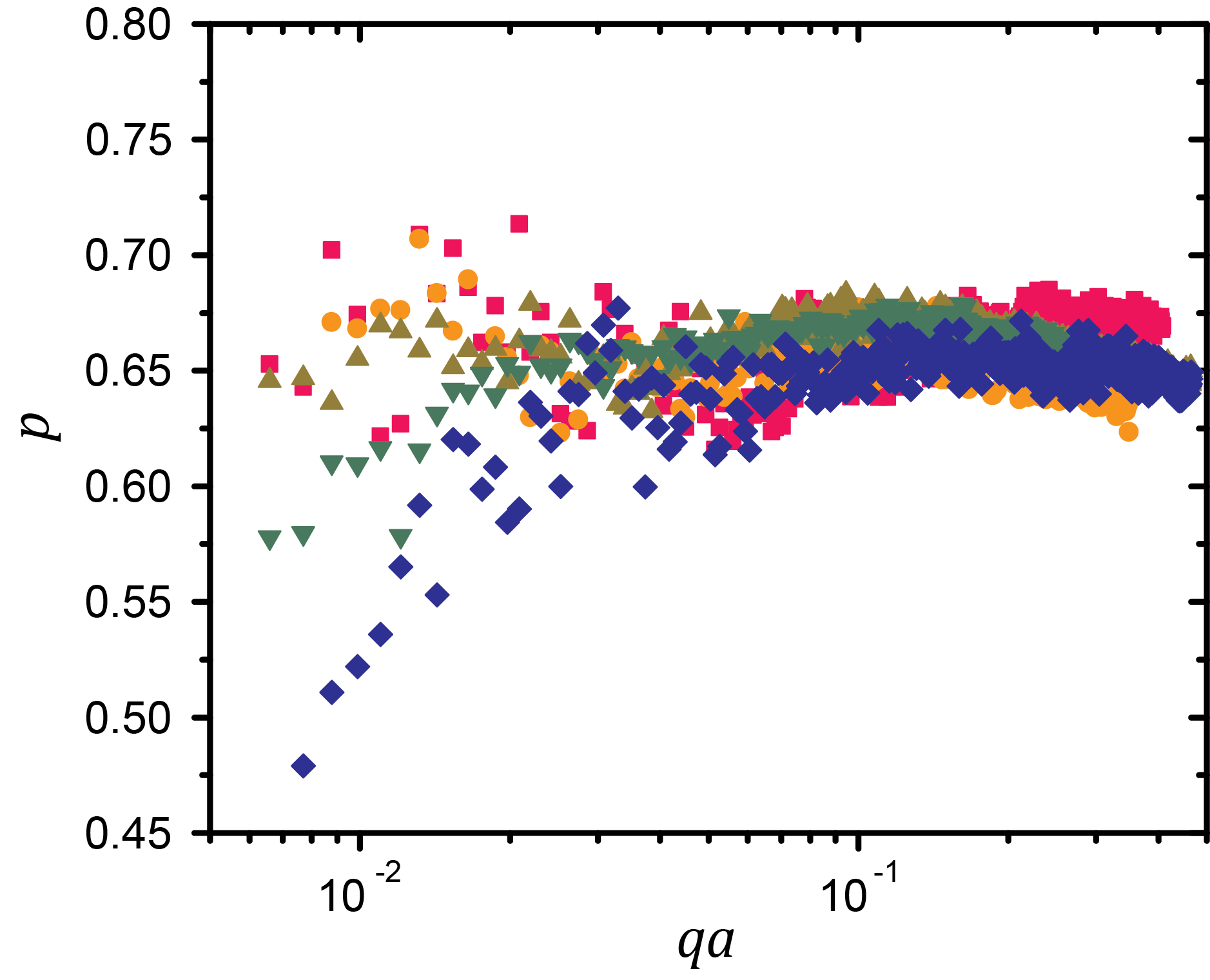}
\caption{\label{supp_fig3} Stretching exponent $p$ of the aged gels in quasi-steady states as a function of the nondimensionalized wave vector $qa$ at $\phi=0.5\%$ ($\square$), $0.8\%$ ($\medcircle$), $1.0\%$ ($\triangle$), $1.5\%$ ($\bigtriangledown$), and $2.0\%$ ($\Diamond$).}
\end{figure}

For all $t$ and $\phi$, the stretching exponent $p(q)$ is largely independent of $q$, as shown in Fig. \ref{supp_fig3}. We find that the level of noise increases for the low-$q$ domain, but we nonetheless assume $q$-independence of $p$, given the robustness of the mean. \par

\subsection{6. Maximum mean squared displacement $\delta^{2}$ of the clusters}

\begin{figure}[h]
\setlength{\abovecaptionskip}{0pt}
\hspace*{-0.03cm}\includegraphics[scale=0.12]{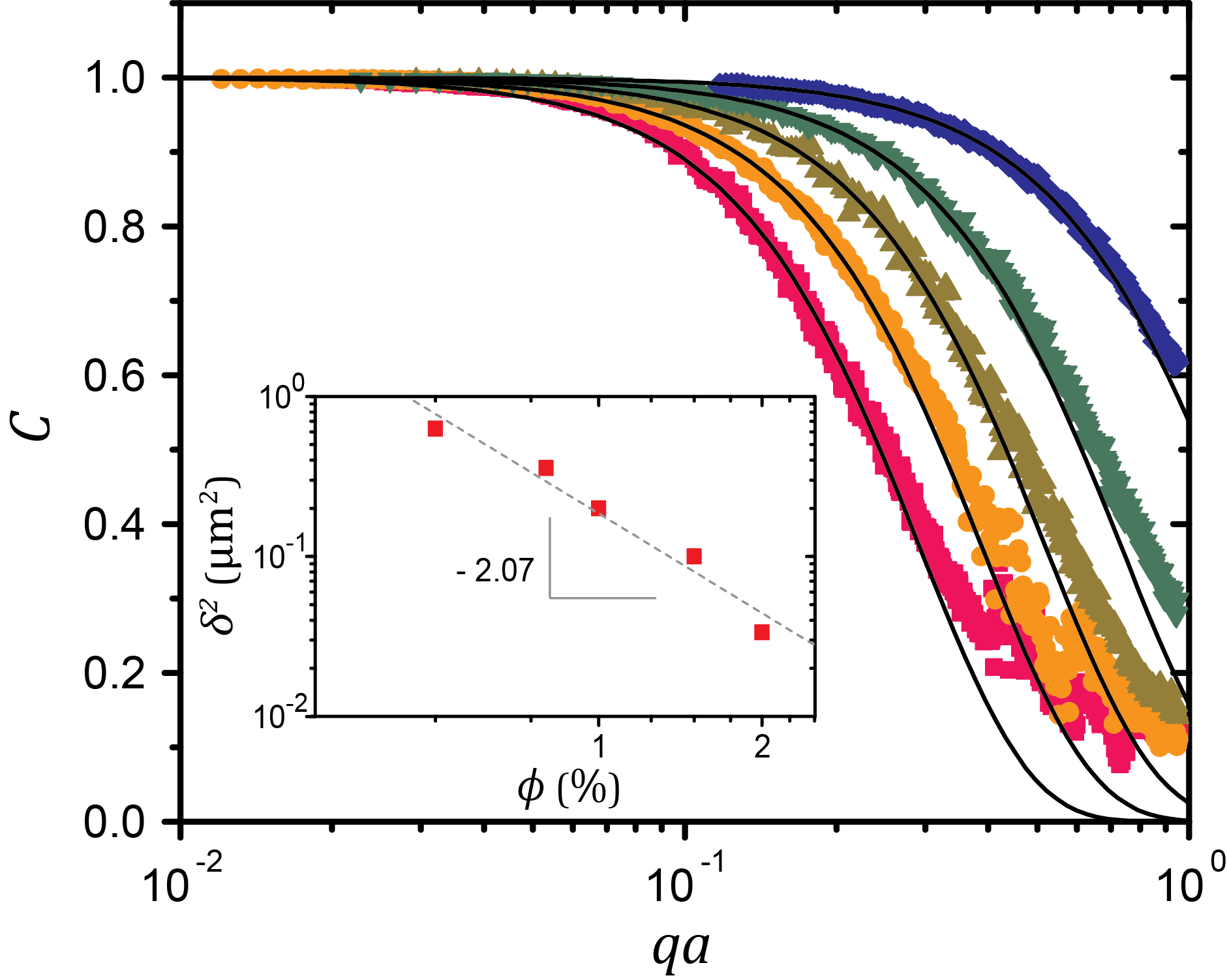}
\caption{\label{supp_fig4} Nonergodicity parameter $C$ of the aged gels in quasi-steady states as a function of the nondimensionalized wave vector $qa$ at $\phi=0.5\%$ ($\square$), $0.8\%$ ($\medcircle$), $1.0\%$ ($\triangle$), $1.5\%$ ($\bigtriangledown$), and $2.0\%$ ($\Diamond$). Black lines are fits of Eq. (\ref{eq4}) to the intermediate-$q$ regimes. Inset: maximum mean squared displacement of the clusters $\delta^{2}$ as a function of $\phi$.}
\end{figure}

We repeat Eq. (5) of the main text here for convenience:
\begin{equation}
f(q,{\Delta}t \rightarrow \infty) = \exp\left(-\frac{q^{2}\delta^{2}}{4}\right) = C(q), \label{eq4}
\end{equation}
where $\delta^{2}$ is the maximum mean squared displacement of the clusters \cite{Krall1998}, and $C(q)$ is the nonergodicity parameter. By fitting Eq. (\ref{eq4}) to $C(q)$ in the intermediate-$q$ domain, where the dynamics is dominated by the clusters, as shown in Fig. \ref{supp_fig4}, we extract the only free parameter $\delta^{2}$. For each $\phi$, Eq. (\ref{eq4}) yields a satisfactory fit until it deviates from $C(q)$ in the high-$q$ domain, where the internal vibrations of fractals determine $\tau(q)$. The resulting values of $\delta^{2}$ as a function of $\phi$ obey a power law with an exponent $-2.07\pm0.04$, as displayed in the inset of Fig. \ref{supp_fig4}. \par

\subsection{7. Direct application of the Krall-Weitz model}

\begin{figure}[h]
\setlength{\abovecaptionskip}{0pt}
\hspace*{-0.03cm}\includegraphics[scale=0.12]{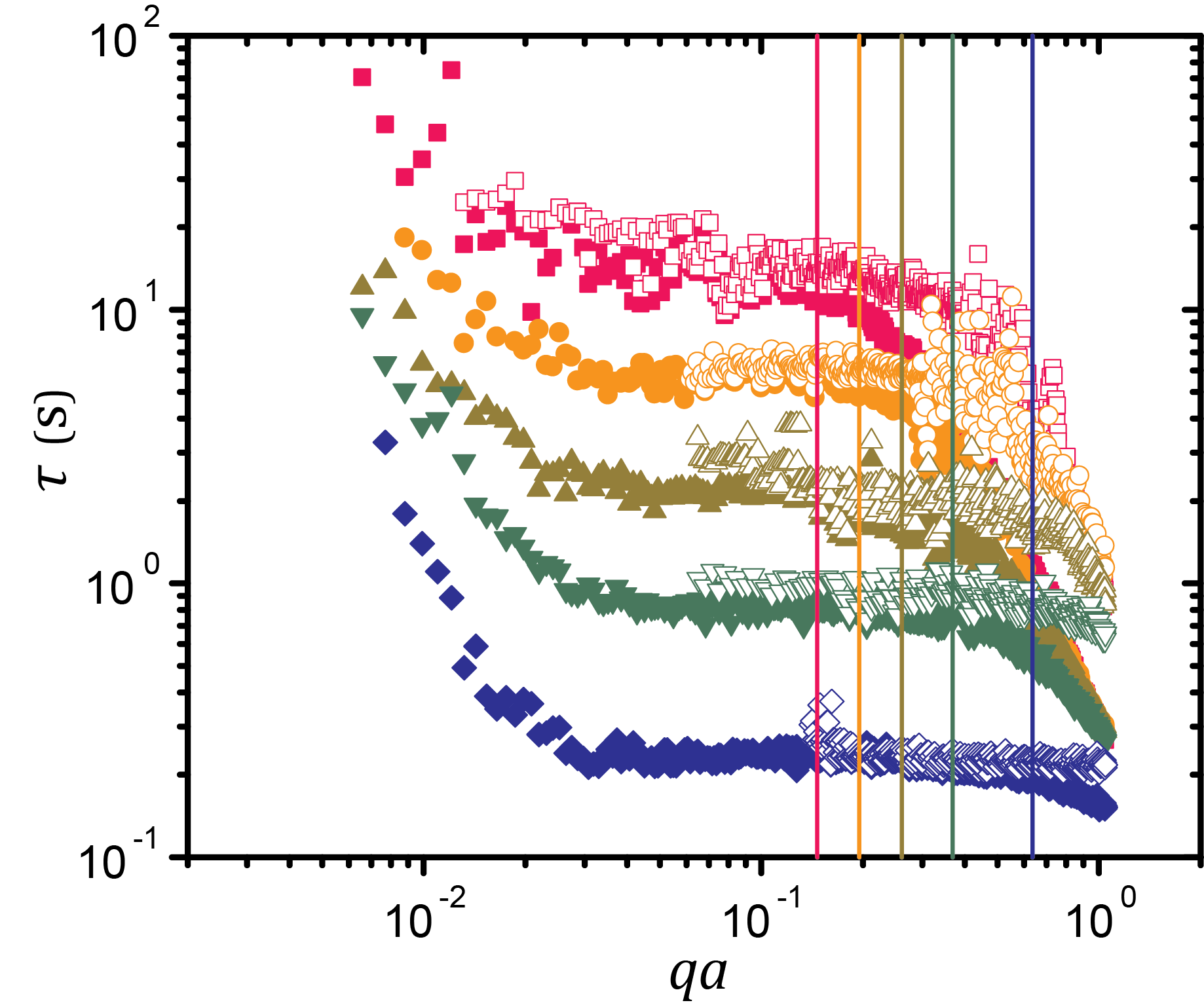}
\caption{\label{supp_fig5} Relaxation time $\tau$ of the aged gels in quasi-steady states as a function of the nondimensionalized wave vector $qa$ at $\phi=0.5\%$ ($\square$), $0.8\%$ ($\medcircle$), $1.0\%$ ($\triangle$), $1.5\%$ ($\bigtriangledown$), and $2.0\%$ ($\Diamond$). Closed symbols represent the values extracted from our model in Eq. (\ref{eq2}) and open symbols from the Krall-Weitz model in Eq. (\ref{eq5}). Vertical lines indicate the values of $qa=a\sqrt{1/\delta^{2}}$, which monotonically increase with $\phi$.}
\end{figure}

To further verify the equivalence of the dynamics in the intermediate-$q$ regime to that of the model reported in Ref. \cite{Krall1998}, hereafter referred to as the Krall-Weitz model, we directly apply the Krall-Weitz model to fit to the image structure function $D(q,{\Delta}t)$ by using Eq. (3) of the main text for the normalized intermediate scattering function, repeated here \cite{Krall1998}:
\begin{equation}
f(q,{\Delta}t) = \exp\left[-\frac{q^{2}\delta^{2}}{4}\left(1-\exp\left[-{\left(\frac{{\Delta}t}{\tau}\right)}^{p}\right]\right)\right]. \label{eq5}
\end{equation}
For both models, we use the same domains of ${\Delta}t$ to fit to $D(q,{\Delta}t)$ for each $q$ and $\phi$. Because of the more complex form of the double exponentials in Eq. (\ref{eq5}), however, we observe that $\tau$ extracted from the Krall-Weitz model is more susceptible to noise, as shown in Fig. \ref{supp_fig5}. We omit $\tau$ of the Krall-Weitz model for lower values of $q$ since the quality of the fits significantly deteriorates. \par

\begin{figure}[h]
\setlength{\abovecaptionskip}{0pt}
\hspace*{-0.03cm}\includegraphics[scale=0.12]{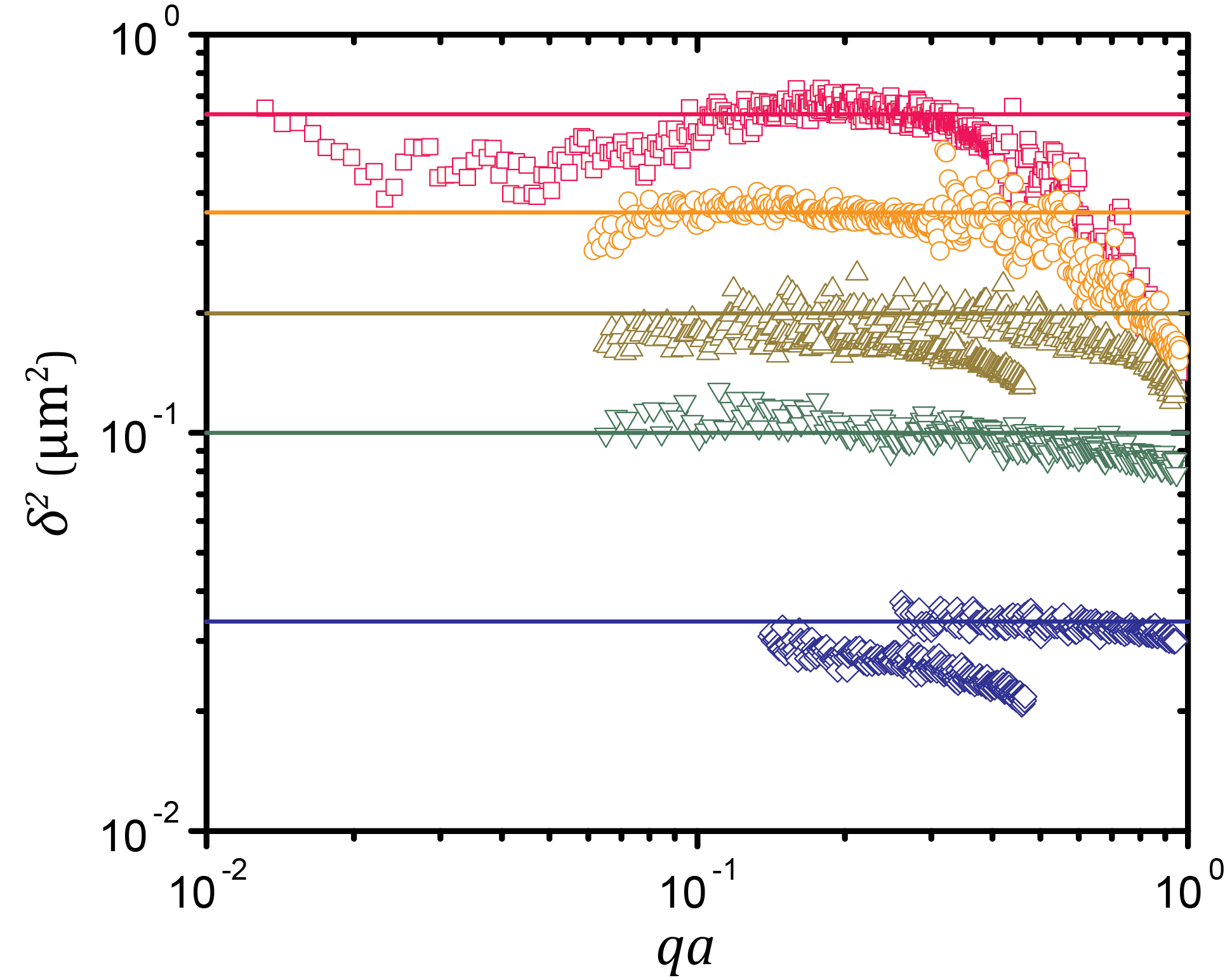}
\caption{\label{supp_fig6} Maximum mean squared displacement $\delta^{2}$ of the aged gels in quasi-steady states as a function of the nondimensionalized wave vector $qa$, extracted from the Krall-Weitz model at $\phi=0.5\%$ ($\square$), $0.8\%$ ($\medcircle$), $1.0\%$ ($\triangle$), $1.5\%$ ($\bigtriangledown$), and $2.0\%$ ($\Diamond$). Horizontal lines indicate the values of $\delta^{2}$ at all $\phi$ obtained from our model as described in Section 6.}
\end{figure}

Eq. (\ref{eq5}) can be simplified to our model in Eq. (\ref{eq2}) for $q^{2}\delta^{2} \ll 1$, and we find that indeed the values of $\tau$ in the two models agree well with each other in the intermediate-$q$ domain, where $q<\sqrt{1/\delta^{2}}$, as displayed in Fig. \ref{supp_fig5}. For the Krall-Weitz model, $\tau$ retains its $q$-independence up to higher values of $q$ than our model, since $q^2$ in Eq. (\ref{eq5}) effectively decreases the characteristic time scale of the outer exponential as $q$ increases. Yet, for the highest accessible $q$ domain, $\tau$ in the Krall-Weitz model also decreases with increasing $q$. Likewise, the values of $\delta^2$ extracted from the two models closely agree with each other in the intermediate-$q$ domain, while in the highest-$q$ domain $\delta^2$ decreases with $q$, as shown in Fig. \ref{supp_fig6}. The $q$-dependence of $\tau$ and $\delta^{2}$ confirms the deviation of the high-$q$ dynamics from the cluster-dominated fluctuations. \par

We finally investigate the $q$-dependence of $f(q,{\Delta}t)$ in the Krall-Weitz model to demonstrate why separate consideration of the characteristic time scale and the nonergodicity, as in Eq. (\ref{eq2}), is necessary to capture both the intermediate-$q$ and the high-$q$ dynamics. In the limit of $({\Delta}t/\tau)^{p}\ll1$, Eq. (\ref{eq5}) can be simplified to
\begin{equation}
f(q,{\Delta}t)\approx\exp\left[-\left(\frac{{\Delta}t}{{\left(4/\delta^{2}\right)}^{\frac{1}{p}}\tau{q}^{-\frac{2}{p}}}\right)^{p}\right]=\exp\left[-\left(\frac{{\Delta}t}{\tau^{*}(q)}\right)^{p} \right], \label{eq6}
\end{equation}
which leads to the scaling of $\tau^{*}\ \sim\ q^{-2/p}$, given that both $\tau$ and $\delta^{2}$ are independent of $q$. Although this result agrees with the $q$-dependence of the relaxation time of the ergodic subdiffusion in the high-$q$ limit within the model by Reuveni et al. \cite{Reuveni2012}, the assumption that $({\Delta}t/\tau)^{p}\ll1$ when $\tau \sim q^{0}$ indicates that this scaling has to hold true at any $q$, if $\tau$ is sufficiently large. Thus, we need to account for the $q$-dependence of the characteristic time scale and that of the nonergodicity separately to properly capture the dynamics in both the intermediate-$q$ domain and the high-$q$ domain. \par 

\subsection{8. Interpretation of elasticity exponent $\beta$}
The value of the elasticity exponent $\beta = 2.50\pm0.12$ obtained from our analysis in the intermediate-$q$ domain reflects considerable structural rearrangements during the network formation. For a percolating chain resistant to bending and stretching, $\kappa \sim N^{-1}R_{\perp}{}^{-2}$, where $N$ is the number of particles and $R_{\perp}$ the radius of gyration of the chain projected onto the plane perpendicular to the line connecting the two ends of the chain \cite{Kantor1984}. For a chain of end-to-end distance $\xi$, $N \sim \xi^{d_{B}}$, where $d_{B}$ denotes the bond dimension, while $R_{\perp} \sim \xi^{\epsilon}$, where $\epsilon$ specifies the extent of isotropy of the chain from $\epsilon = 0$ for a perfectly straight chain to $\epsilon = 1$ for an isotropic chain \cite{DeRooij1994}. Consequently, the following scaling relation holds \cite{DeRooij1994, Romer2014}:
\begin{equation}
\kappa\ \sim\ \xi^{-(2\epsilon+d_{B})}\ \sim\ \xi^{-\beta}. \label{eq7} 
\end{equation}
The elasticity of a network of fractal aggregates, therefore, is largely conferred by the geometry of the force-bearing backbones. For an ideal network of compactly packed fractal clusters, $\epsilon\approx1$.  Substituting $\beta = 2.50\pm0.12$ and the representative value of $d_{B} = 1.1$ \cite{Meakin1984} into Eq. (\ref{eq7}), however, we find $\epsilon = 0.70\pm0.06$, implying significant anisotropy of the stress-bearing strands. In Fig. 4(c) of the main text, we confirm the prevalence of voids present among the fractal clusters, caused by structural rearrangements that deflect the gel backbones to be anisotropic.\par

\bibliographystyle{apsrev4-1}
\bibliography{DDM_paper_Aug_19_SM_mod.bib}